\documentclass[aps,twocolumn,amsfonts,amssymb,longbibliography,superscriptaddress]{revtex4-2}
\usepackage{graphicx}
\usepackage{multirow}
\usepackage{booktabs}
\usepackage{epstopdf} 
\usepackage{dcolumn}
\usepackage{bm}
\usepackage{bbm,bbold}
\usepackage{color}
\usepackage{xcolor, soul}
\sethlcolor{green}
\usepackage{amsfonts}
\usepackage{amsmath}
\usepackage{mdframed}
\usepackage[normalem]{ulem}
\usepackage{mathrsfs}   
\usepackage[none]{hyphenat}
\usepackage{subfigure}
\usepackage{physics}
\usepackage[colorlinks=true,citecolor=blue]{hyperref}
\hypersetup{colorlinks=true,citecolor=blue,linkcolor=blue,urlcolor=blue}

\usepackage{booktabs}
\usepackage{graphicx}
\usepackage{multirow}
\usepackage{makecell}
\usepackage{booktabs}
\usepackage{pifont}
\usepackage{tabularx}

\usepackage[colorlinks=true,citecolor=blue]{hyperref}
\begin{document}
	\title{Interband response in spin–orbit coupled nodal line semimetals}	
	\author{Vivek Pandey}
        \email{vivek_pandey@srmap.edu.in}
	\affiliation{Department of Physics, School of Engineering and Sciences, SRM University AP, Amaravati, 522240, India}
    \author{Monu}
        \email{monu_bijendra@srmap.edu.in}
	\affiliation{Department of Physics, School of Engineering and Sciences, SRM University AP, Amaravati, 522240, India}
 	\author{Pankaj Bhalla}
	\email{pankaj.b@srmap.edu.in}
 \affiliation{Department of Physics, School of Engineering and Sciences, SRM University AP, Amaravati, 522240, India}
\affiliation{Centre for Computational and Integrative Sciences, SRM University AP, Amaravati, 522240, India}
    
\date{\today}

\begin{abstract}
This study investigates the interband conductivity for nodal line semimetals (NLSMs) in the presence of spin-orbit coupling (SOC), where the disorder reshapes the transport properties. The SOC breaks spin degeneracy, thus fundamentally altering the band dispersion and enabling multiple interband transport channels. Using a quantum kinetic framework, we analyze the interband conductivity originating from disorder-driven (extrinsic) and field-driven (intrinsic) mechanisms. We find that the interband response shows an anisotropic nature due to disorder driven counterparts. Additionally, our predictions show a tunable prominent transition peak arising from non-Pauli-blocked states that can be controlled via band parameters as well as external stimuli. To have an experimental relevance, we provide a numerical estimation for the interband response of TaAs using density functional theory estimated parameters. These results suggest the investigation of disorder-enabled signatures in spin systems. 

\end{abstract}

\maketitle
\section{Introduction}
\begin{figure}[t]
    \centering
    \includegraphics[width=7.8 cm]{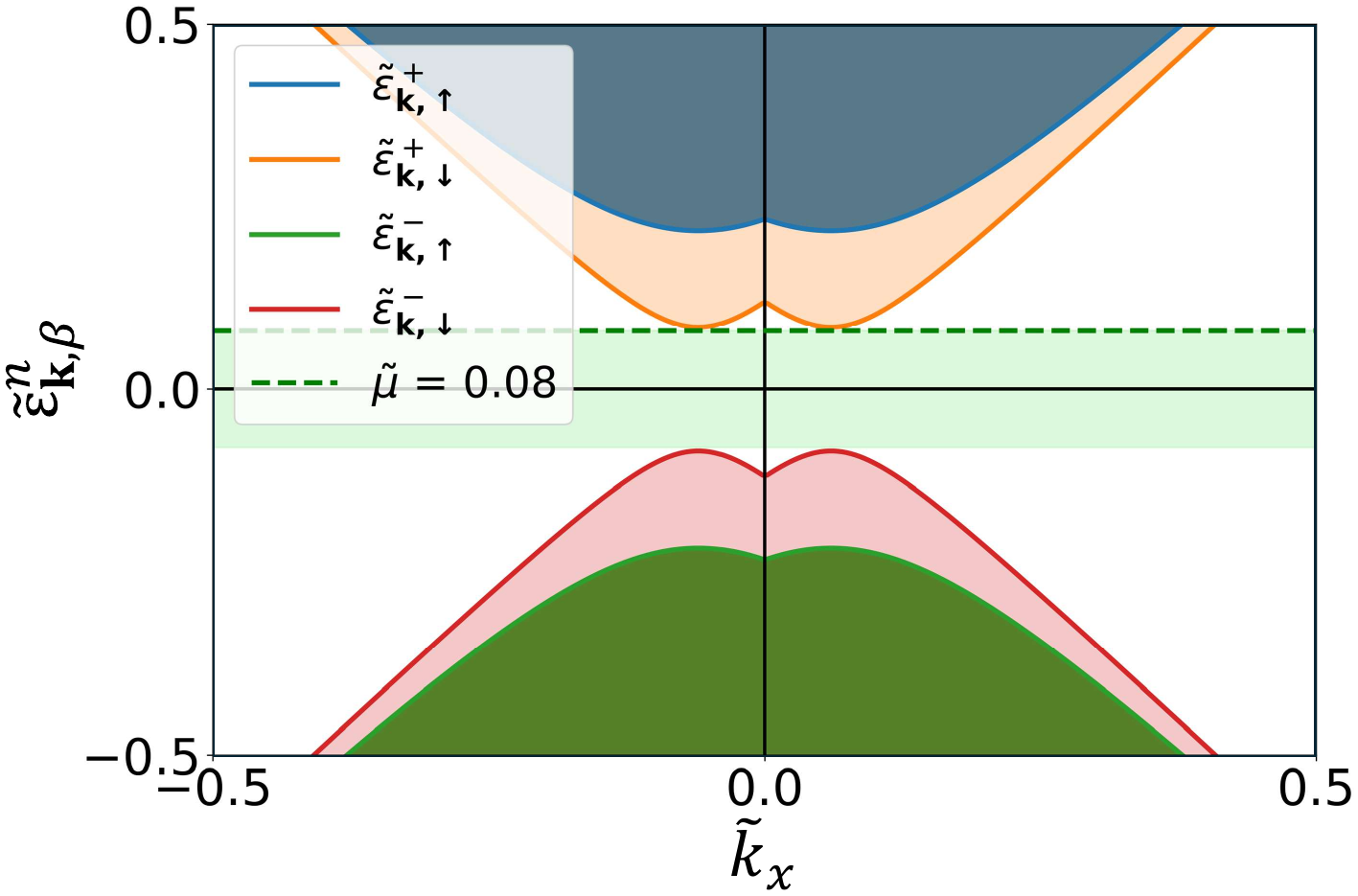}
        \caption{Depicts the two-dimensional (2D) band dispersion of a nodal line semimetal with spin-orbit coupling (SOC) in the regime $k_0>m'$. SOC lifts the spin degeneracy, leading to spin-split bands. The dotted line indicates the chemical potential, and the shaded green region denotes the SOC-induced gap. Here, $k_0$ represents the nodal ring radius, $m'$ is the parameter associated with the SOC-related term. Dispersion is plotted at $\tilde{k}_y=0, \tilde{k}_z= 0.1$, band index $n=\pm1$ and spin index $\beta=\pm 1$.} 
    \label{fig:1D_dispersion}
\end{figure}

The study of transport in topological semimetals (TSMs) remains a central topic in condensed matter physics, due to their non-trivial electronic properties arising from symmetry-enforced band crossings. Depending on the key characteristic of band crossing, such as its degeneracy and dimension difference (i.e., band crossing in a point or a line)~\cite{Gao_armr2019}, TSMs can be broadly classified into Dirac (DSMs)~\cite{Armitage_rmp2018, Young_prb2012,liu_nm2014, Borisenko_prl2014, yi_sr2014,  Wang_prb2013, liu_science2014}, Weyl (WSMs)~\cite{Armitage_rmp2018, Yan_arcmp2017, Lv_prx2015, soluyanov_n2015, Hasan_arcmp2017, Weng_prx2015, xu_nc2016, Belopolski_prl2016, Xu_sa2015, Zyuzin_prb2012} and nodal line semimetals (NLSMs)~\cite{Fang_cpb2016, Fang_prb2015, Burkov_prb2011, Shuo_apx2018, Bian_prb2016, Xie_APLM2015, Kim_prl2015, Ekahana_njp2017, Xu_prb2017, Takane_prb2017, Chen_prb2017, Wang_prb2017, chang_arxiv2025, Zhu_cpl2024}. Here, DSMs are fourfold degenerate, where the conduction and valence bands touch at a single point, breaking any of the underlying symmetries (i.e., either time reversal ($\mathcal{T}$) or space inversion ($\mathcal{P}$)) can generate WSMs, where bands touch each other in a pair of two-point degenerate Weyl points. In NLSMs, band touching occurs either in a line or a loop, making it more robust and interesting for further study.
%
Further, the presence of spin-orbit coupling (SOC) in TSMs fundamentally modifies the electronic structure and gives rise to a variety of emergent phenomena rooted in symmetry breaking, such as the loss of SU(2), inversion, or time-reversal symmetry~\cite{Fang_cpb2016, Fang_prb2015}. 
In Dirac semimetals, SOC stabilizes Dirac points even in the absence of inversion symmetry, leading to non-zero Berry curvature and symmetry-protected edge states~\cite{Jin_prl2020}. A prominent example among DSMs is Na$_3$Bi, where SOC not only modifies the band structure but also influences carrier dynamics~\cite{tancogne_npj2022}. 
In contrast, SOC transforms the NLSMs into various phases such as multiple Weyl points, Dirac points, or fully gapped WSMs via lifting the spin degeneracy of the system~\cite{Fang_cpb2016, bian_nc2016, Wang_sc2021}, as depicted in Fig.~\ref{fig:1D_dispersion}. %
%
%
Such evolutions have been observed in the first-principle calculations of TaAs, TaP, NbAs, NbP, and SrIrO$_3$~\cite{Weng_prx2015, huang_nc2015, Pandey_jpcm2023, Carter_prb2012, fang_np2016}. 

In TSMs, the transport response mainly originates from two processes, namely intraband and interband. The intraband part of the response arises from carrier motion within the same band, follows a Drude-like behavior, and usually dominates at low frequencies~\cite{Kandel_jpcm2024, Barati_prb2017}. However, the interband part of the response coming from the carrier motion between two distinct bands becomes significant in the mid and high frequency regimes~\cite{Barati_prb2017, Pandey_prb2024, Pandey_pssrrl2025}.

Recent studies reveal that in the field-driven (intrinsic) interband part of the conductivity in spin-degenerate 3D NLSMs shows a nontrivial dependence on chemical potential ($\mu$), exhibiting linear and quadratic dependence in different $\mu$ regimes~\cite{Burkov_prb2011, Barati_prb2017, Mukherjee_prb2017}. 
In the 2D system, the optical conductivity is solely intraband with Drude-like behavior as the interband part vanishes in the absence of momentum-dependent pseudospin textures in the Hamiltonian~\cite{Barati_prb2017}.
Inclusion of disorder further enriches the picture by showing the dominant nature of extrinsic contributions over others to the DC and optical conductivity of 3D systems~\cite{Pandey_prb2024, Pandey_pssrrl2025}. %
Such disorder-induced contributions show significant impact on the response of 3D NLSMs and are sensitive to the symmetry-breaking a gap-induced term in the band dispersion~\cite{Pandey_prb2024, Pandey_pssrrl2025}. On the other hand, for 2D NLSMs, the extrinsic contribution vanishes, leaving the transport insensitive to scattering-driven interband mechanism.~\cite{Barati_prb2017}. Here, we investigate the response for a non-degenerate NLSM system incorporating SOC. The presented framework reveals several unique features as follows: 
%
    First, the SOC breaks the spin degeneracy, causing two bands structure to be transformed into four band structure. This band non-degeneracy leads to new interband transition channels, i.e., spin-down valence band to spin-down conduction band and spin-up valence band to spin-up conduction band, unlike the previous works~\cite{Barati_prb2017, Pandey_prb2024, Pandey_pssrrl2025}, where the interband transition is limited to valence-conduction band transition. Further, these spin-resolved interband transitions 
    yields resonance features in the longitudinal conductivity that are tunable by the nodal ring radius, the SOC term, the chemical potential, and the frequency.
    %
    Second, the SOC breaks the SU(2) symmetry and causes the phase transition from nodal line to Weyl-like or fully gapped phases, depending on nodal ring radius and SOC term. This topological phase transition gives significant impact on the conductivity, which is absent in a spin-degenerate system.
    Third, the total interband conductivity shows an anisotropic directional dependence, which mainly arises from the scattering-driven interband mechanism, while the field-driven mechanism is largely isotropic in nature. This anisotropic nature can serve as an experimental tool for accessing the SOC-induced transport response in topological systems.
    In addition, the extrinsic response shows a tunable characteristic transition peak with nodal ring radius and SOC-related parameters, as well as external probes such as frequency and chemical potential. However, the intrinsic component shows significant tunability with the external parameters.
    Furthermore, the previous studies~\cite{Pandey_prb2024, Pandey_pssrrl2025} mainly emphasized on the disorder-induced interband transport within the same band topology. On the contrary, the interplay of spin splitting, symmetry breaking, and topology adds new controllable features to the disorder induced interband transport.
%
%
%
%

%
%

In this work, we present a comprehensive and unified theoretical framework to unveil the disorder-induced interband transport response in NLSMs with SOC, using a quantum kinetic theory. By incorporating the material-specific parameters for TaAs, we provide a bridge between our theoretical predictions and experimentally relevant transport signatures. Moreover, the presented predictions will be useful for spintronic devices and topological transistors.
\begin{figure*}[t]
    \centering
    \includegraphics[width=15 cm]{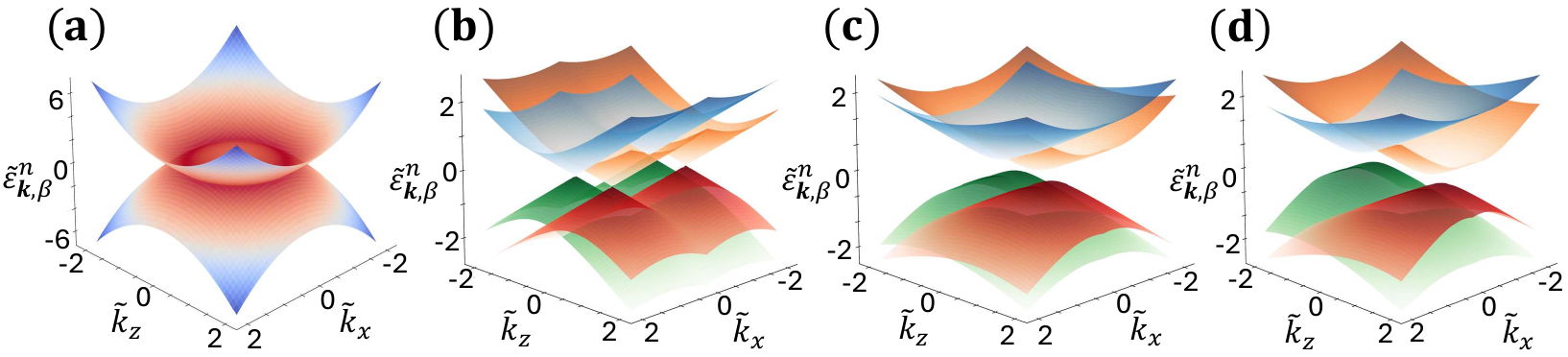}
    \caption{Shows the three-dimensional (3D) band dispersion of the NLSM with and without SOC. (a): dispersion without the SOC case, where the nodal ring is symmetry-protected. (b)-(d): depict the evolution of the 3D band structure in the presence of SOC for different regimes of the parameters $k_0$ and $m'$. (b): $k_0>m'$ the system shows a Weyl-like behavior characterized by a pair of Weyl nodes. In the regimes $k_0=m'$ and $k_0<m'$, shown in panels (c) and (d), the Weyl nodes annihilate, and the system becomes gapped. Here, $\tilde{k}_y=0$, $n$ and $\beta$ $=\pm 1$.} 
    \label{fig:3D_dispersion}
\end{figure*}
\section{Model and Theoretical Setup}
\subsection{Model}\label{subsec:model}
The four-band low-energy model Hamiltonian for a nodal line semimetal in the presence of SOC based on first principle calculation is~\cite{Fang_cpb2016, Weng_prx2015, huang_nc2015} 
\begin{align}\nonumber
    &\mathcal{H}(\bm k) =\\& \varepsilon_0 [(\tilde{\mathcal{K}} - 1)\sigma_z \otimes s_0 + \gamma \tilde{k}_z \sigma_y \otimes s_0 - \tilde{m}^\prime \sigma_y \otimes s_x +\gamma' \tilde{k}_x \sigma_x \otimes s_x].
\end{align}
Here $\tilde{\mathcal{K}}=\mathcal{K}/k_0$, where $\mathcal{K}=\sqrt{k_x^2+k_y^2}$ and $k_0$ is the radius of the nodal ring. The $\tilde{k}_i= k_i/k_0$, where $k_i$ refers to the wave vector linked with the electron in the $i$th direction having $i\equiv x, y,z$. Further, $\sigma_i$ and $s_i$ represent the Pauli matrices in the orbital and spin basis, respectively. Additionally, $\varepsilon_0 = \hbar^2 k_0^2/2m_e$ represents the energy associated with the nodal ring, where $m_e$ is the mass of the electron and $\hbar$ is the reduced Planck constant. $\gamma = 2m_ev_z/\hbar k_0$, $\gamma'=2m_ev_x/\hbar k_0$ and $\tilde{m}'= 2m_e v_y m'/\hbar k_0^2$.  
In the above Hamiltonian, the last two terms $- \tilde{m}^\prime \sigma_y \otimes s_x + \gamma' \tilde{k}_x \sigma_x \otimes s_x$ correspond to the spin-orbit coupling terms, where $m'$, having dimensions of a wave vector.
%
%
The presence of such SOC terms lifts the spin degeneracy, leading to finite band repulsion between opposite spin states at finite momentum, which makes the nodal line unstable. Further, the addition of SOC breaks SU(2) symmetry of the system, thus evolving the nodal line into various scenarios such as Weyl point pairs, or a completely gapped system~\cite{Fang_cpb2016, Fang_prb2015, Weng_prb2015}.
However, in the absence of SOC terms, the system preserves combined space inversion and time reversal ($\mathcal{PT}$) symmetry, mirror symmetry and SU(2) spin rotation symmetry~\cite{Barati_prb2017, Pandey_prb2024, Weng_prb2015, Wang_prb2021}.
The band dispersion corresponding to the Hamiltonian is
\begin{align}
    \tilde{\varepsilon}^n_{\bm k, \beta} = n \sqrt{(\tilde{\mathcal{K}} -1)^2 + (\gamma\tilde{k}_z+\beta \tilde{m}^\prime)^2 + (\gamma'\tilde{k}_x)^2},
\end{align}
where $n=\pm 1$ and $\beta = \pm 1$ refer to the band and spin indices, respectively.
The dispersion is displayed in Fig.~\ref{fig:3D_dispersion} for different cases. First, in the absence of SOC, the nodal ring is protected and the bands are degenerate as shown in Fig.~\ref{fig:3D_dispersion}(a). Upon introducing finite SOC,  the band degeneracy is lifted and anisotropy in the band structure dispersion adds up due to the finite $m'$ term, which is particularly evident in the $k_z$ direction. Based on the competition between $k_0$ and $m'$, there can be distinct scenarios as shown in Figs.~\ref{fig:3D_dispersion} (b)-(d). 
For $k_0>m'$, the nodal ring splits into distinct Weyl nodes, indicating a topological phase transition from a nodal line to a Weyl semimetal phase. Furthermore, in the cases of $k_0=m'$ and  $k_0<m'$, the Weyl node annihilates and the system becomes gapped.

\subsection{Theoretical Setup}
The setup follows the quantum kinetic approach for the single-particle density matrix $\rho$ within the band basis representation~\cite{pottier_OUP, Culcer_prb2017, Bhalla_prb2023},
\begin{equation} \label{eqn:QLE}
    \frac{\partial \rho^{np}}{\partial t}+\frac{i}{\hbar}[\mathcal{H}_0,\rho]^{np} +\mathcal{J}^{np}[\rho] = -\frac{i}{\hbar}[\mathcal{H}_E,\rho]^{np}. 
\end{equation}
Here $[\cdot,\cdot]$ represents the commutator bracket, $\mathcal{H}_0$ is the unperturbed part of the Hamiltonian.
The field part $\mathcal{H}_{\bm E} =e\bm{E}\cdot\hat{\bm{r}}$, stems from the interaction of the external electric field $\bm{E}$, which is taken as time-dependent and spatially homogeneous, $\hat{\bm{r}}$ and $e$ represent a position vector and charge associated with the electron, respectively. 
The disorder part $U$ contained in $\mathcal{J}^{np}[\rho]=i/\hbar[U,\rho]^{np}$ represents the scattering contribution that arises from the disorder term. In the present study, we consider a weak disorder potential for which the average of the disorder matrix elements is zero. 
However, the multiple averages of the elements of the disorder matrix in a unit volume are $\langle U(\bm{r})U(\bm{r}')\rangle=U_0^2\delta(\bm{r}-\bm{r}')$~\cite{Culcer_prb2017, Yang_prb2011}, here $U_0$ incorporates the strength of the disorder potential. 

To understand the band dynamics, the density matrix can be broken as: $\rho^{np} = \rho_{0,\bm{k}}^{nn} + \mathcal{N}_{\bm E}^{nn} + S_{\bm E}^{np} $. 
Here, $\mathcal{N}_{\bm E}^{nn}$ and $S_{\bm E}^{np}$ are the intraband and interband field correction terms, respectively. Furthermore, $\rho_{0}^{nn}=f^0_{\bm{k}}=\big[1+e^{\beta(\varepsilon_{\bm{k}, \beta}^n-\mu)}\big]^{-1}$ is the equilibrium Fermi-Dirac distribution function having $\beta = [k_BT]^{-1}$, $k_B$ representing the Boltzmann constant, $T$ and $\mu$ are the absolute electronic temperature and the chemical potential, respectively. Here, $\varepsilon^n_{\bm k, \beta}$ is the eigenenergy associated with the $n^{\text{th}}$ band. 
Further, the solution of Eq.~\eqref{eqn:QLE} incorporating the intraband and interband contributions can be obtained by replacing $\rho^{np}$ with $\mathcal{N}_{\bm E}^{nn}$ and $\mathcal{S}_{\bm E}^{np}$ respectively. First, the intraband part yields $ \mathcal{N}_{\bm E}^{nn}=e\bm{E}\space \partial _{\bm{k}}f^0_{\bm{k}}/(g_{\bm k}+i\hbar\omega)$, 
where $g_{\bm k} = \hbar/ \tau^n_{\bm k}$ is the scattering energy scale, $\partial_{\bm k}\equiv \partial/\partial \bm k$ is the partial derivative with respect to the wave vector, $\tau^n_{\bm k}$ is defined as $\hbar/\tau_{\bm k}^n=n_i U_0^2 \sum_{\bm{k'}} \delta(\varepsilon_{\bm {k}',\beta}^n- \varepsilon_{\bm {k}, \beta}^n)$ having $n_i$ as the impurity density. For the interband part, we have $S_{\bm E}^{np}= \frac{{D_{\bm E,{\bm{k}}}^{np}} - \mathcal{J}_{\bm{k}}^{np}[\mathcal{N}_{\bm E,{\bm{k}}}]}{\eta \,+\space i\space (\omega^{np}+ \hbar \omega)}$~\cite{Pandey_prb2024, Pandey_njp2025}.
Here $\omega^{np}= {\varepsilon}^{n}_{\bm k,\beta}- {\varepsilon}^{p}_{\bm k, \beta}$ is the energy difference between the bands, $\eta$ is an infinitesimally small factor and ${D_{\bm E,{\bm{k}}}^{np}}$ is the field-driven (intrinsic) part which is defined like $ D_{\bm E,\bm{k}}^{np} = \langle n|[\mathcal{H}_{\bm E},\rho]\ket{p} = e\bm{E}/\hbar.\left(\partial_{\bm {k}} \rho^{np}-i[\mathcal{R}_{\bm{k}},\rho]^{np}\right)$~\cite{Bhalla_prb2023, Nagaosa_AM2017}, the Berry connection in the $\bm{k}$ space is denoted by $\mathcal{R}_{\bm{k}}^{np}=\langle{u_{\bm{k}}^n}\ket{i\partial_{\bm{k}}u_{\bm{k}}^p}$, where   $|u_{{\bm k}}^{n} \rangle$ is the eigenfunction arriving from the system's Hamiltonian. The $\mathcal{J}_{\bm{k}}^{np}[\mathcal{N}_{\bm E}]$ is the scattering-driven (extrinsic) term which is treated within the first order Born approximation and yields the leading order $\propto U_0^2$~\cite{Culcer_prb2017}.

\section{Interband conductivity in NLSMs}
In general, conductivity can be extracted using the relation $\bm j =\sigma \bm E= \text{Tr}[{\bm v}\rho]$, where $\bm j$ is the current density and $\sigma$ is the conductivity, consisting of intraband and interband parts, i.e., $\sigma^{\text{Total}} = \sigma^{\text{Intra}}+ \sigma^{\text{Inter}}$, where the interband contribution has intrinsic and extrinsic parts, i.e., $\sigma^{\text{Inter}}=\sigma^{\text{Int}} + \sigma^{\text{Ext}}$. The present analysis investigates the interband response, which dominates at $\omega\tau \gg 1$ in comparison to the intraband response, having a major contribution at $\omega\tau \ll 1$. Here, $\bm v$ is the velocity associated with the electron and in the band basis is defined as $ \space \tilde{v}^{pn}_i=\space\delta_{pn}\space\partial_{\tilde{k}_i}\tilde{\varepsilon}_{\bm{{k}},\beta}^n+i\space\mathcal{\tilde{R}}^{pn}_{i}\space\tilde{\omega}^{pn}$, where $\tilde{v}_i=\hbar  {k}_0 v_i/\varepsilon_0 $, $\tilde{\omega}^ {pn}= \omega^{pn}/\varepsilon_0$ and $\mathcal{\tilde{R}}_i=\mathcal{\tilde{R}}_{k_i}$. Following the theoretical setup for the density matrix as discussed in the previous section, the interband conductivity can be expressed in the form
\begin{align}
\label{eqn:total_zz}
\sigma_{ii}^{\text{Inter}}=e \sum_{n\neq p}\sum_{{\bm k}}  \frac{ \mathcal{\tilde{R}}_{i}^{pn} \tilde{\omega}^{pn}}{\eta+\space i \space ( \tilde{\omega}^{np} + \tilde{\omega})} \Big(\underbrace{\frac{e \mathcal{\tilde{R}}_{i}^{np}\tilde{F}^{np}}{\hbar} }_{\text{Intrinsic}} - \underbrace{\frac{i\mathcal{J}_{\tilde{{k}}_i}^{np}[\mathcal{N}_{\bm E}]}{E_i}}_{\text{Extrinsic}}\Big).
\end{align}
Here, the detailed derivation of the above expression is provided in Appendix~\ref {Appendix:A}. Further, $\tilde{\omega}=\omega/\varepsilon_0$ and $\tilde{F}^{np}=f^0{(\tilde{\varepsilon}_{\bm{k},\beta}^n)}-f^0{(\tilde{\varepsilon}_{\bm{k},\beta}^p)}$ represents the difference in equilibrium occupation probabilities of the two bands and governs the possibility of allowed interband transitions through the Pauli exclusion principle. %
\begin{figure*}[t]
    \centering
    \includegraphics[width=17cm]{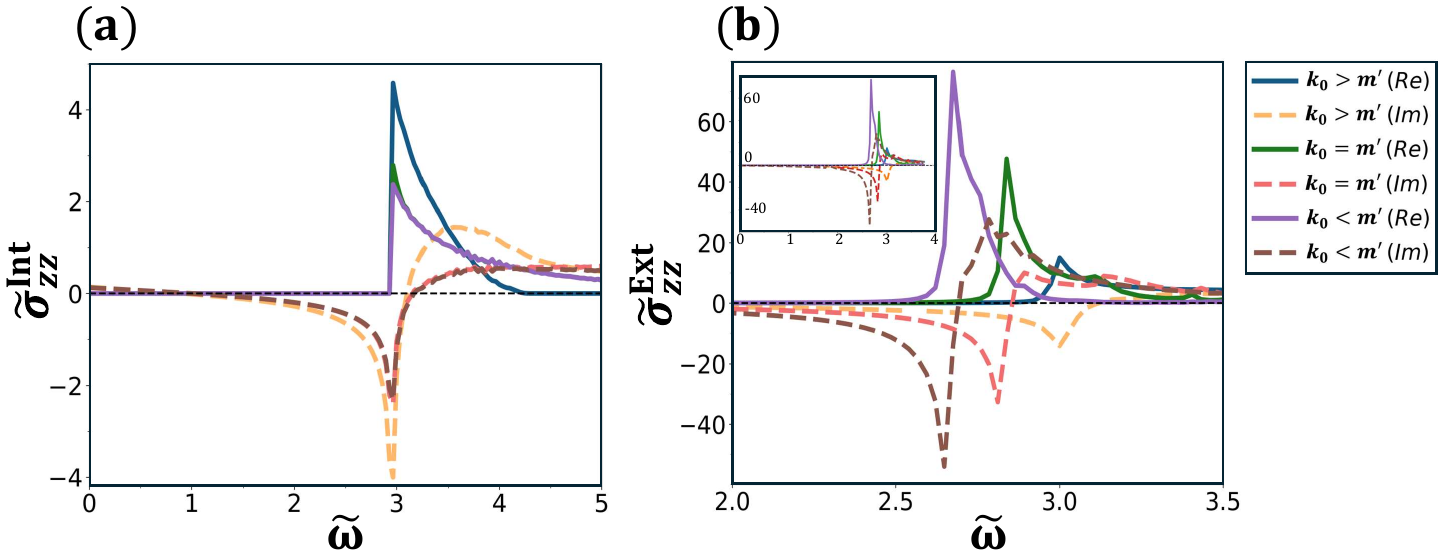}
    \caption{ (a) and (b) illustrate the real part (solid line) and the imaginary part (dotted line) of the normalized intrinsic $\tilde{\sigma}_{zz}^{\text{Int}}=\sigma_{zz}^{\text{Int}}/\sigma_0$ and extrinsic $\tilde{\sigma}_{zz}^{\text{Ext}}=\sigma_{zz}^{\text{Ext}}/\sigma_0$ interband components, respectively, as a function of the normalized frequency ($\tilde{\omega}$). Here, $\sigma_0=e^2/\hbar(2\pi)^2$. The results are plotted for distinct limits of nodal ring radius $k_0$ and term $m'$ associated with the SOC.   The inset in panel (b) shows the variation of extrinsic contribution in the complete range of frequency $\tilde{\omega}=0$ to $4$. All plots are obtained using Eq.~\eqref{eqn:total_zz} and the model Hamiltonian for NLSM with TaAs material parameters and keeping the chemical potential $\tilde{\mu}=1.5$.} 
    \label{fig:Int_Ext_m_mp}
\end{figure*}
Here, the scattering term is taken within the weak disorder limit, i.e., $\mu \tau/\hbar \gg 1$ or $k_F l \gg 1$ and treated using the first-order Born approximation.  
The scattering term $\mathcal{J}_{\bm k}^{np}[\mathcal{N}_{\bm E}]$ in Eq.~\eqref{eqn:total_zz}, shows explicit dependence on $\partial f^0{(\tilde{\varepsilon}_{\bm{k}, \beta}^n)}/\partial \tilde{\varepsilon}^{n}_{\bm{k}, \beta}$ embedded in the $\mathcal{N}_{\bm E}$, further, within the low temperature regime it approximates as $-\delta(\tilde{\varepsilon}^n_{\bm k, \beta}-\tilde{\mu})$, where $\tilde{\mu} = \mu/\varepsilon_0$. 
Furthermore, in the case of a four-band model for NLSMs with broken spin degeneracy due to SOC, the distinct bands can lead to intrinsic and extrinsic contributions. Specifically, the non-zero interband response only arises from the $\tilde{\mathcal{R}}_i^{02}$, $\tilde{\mathcal{R}}_i^{20}$, $\tilde{\mathcal{R}}_i^{13}$ and $\tilde{\mathcal{R}}_i^{31}$ part of the interband Berry connection. Hence, causing the intrinsic contribution to emerge from the combination of these non-zero parts of the Berry connection. Further, in the extrinsic part, only non-zero contributions arise from four components such as $\mathcal{J}_{\bm{k}}^{02}[\mathcal{N}_{\bm E}]$, $\mathcal{J}_{\bm{k}}^{20}[\mathcal{N}_{\bm E}]$, $\mathcal{J}_{\bm{k}}^{13}[\mathcal{N}_{\bm E}]$ and $\mathcal{J}_{\bm{k}}^{31}[\mathcal{N}_{\bm E}]$. The band index 0,1,2,3, correspond to the bands, namely the spin-down valence band ($\downarrow,-$), spin-up valence band ($\uparrow,-$), spin-down conduction band ($\downarrow,+$), and the spin-up conduction band ($\uparrow,+$), respectively. Notably, the only possible contribution arises from the spin-up to spin-up and the spin-down to spin-down transitions.
\begin{figure*}[t]
    \centering
    \includegraphics[width=15cm]{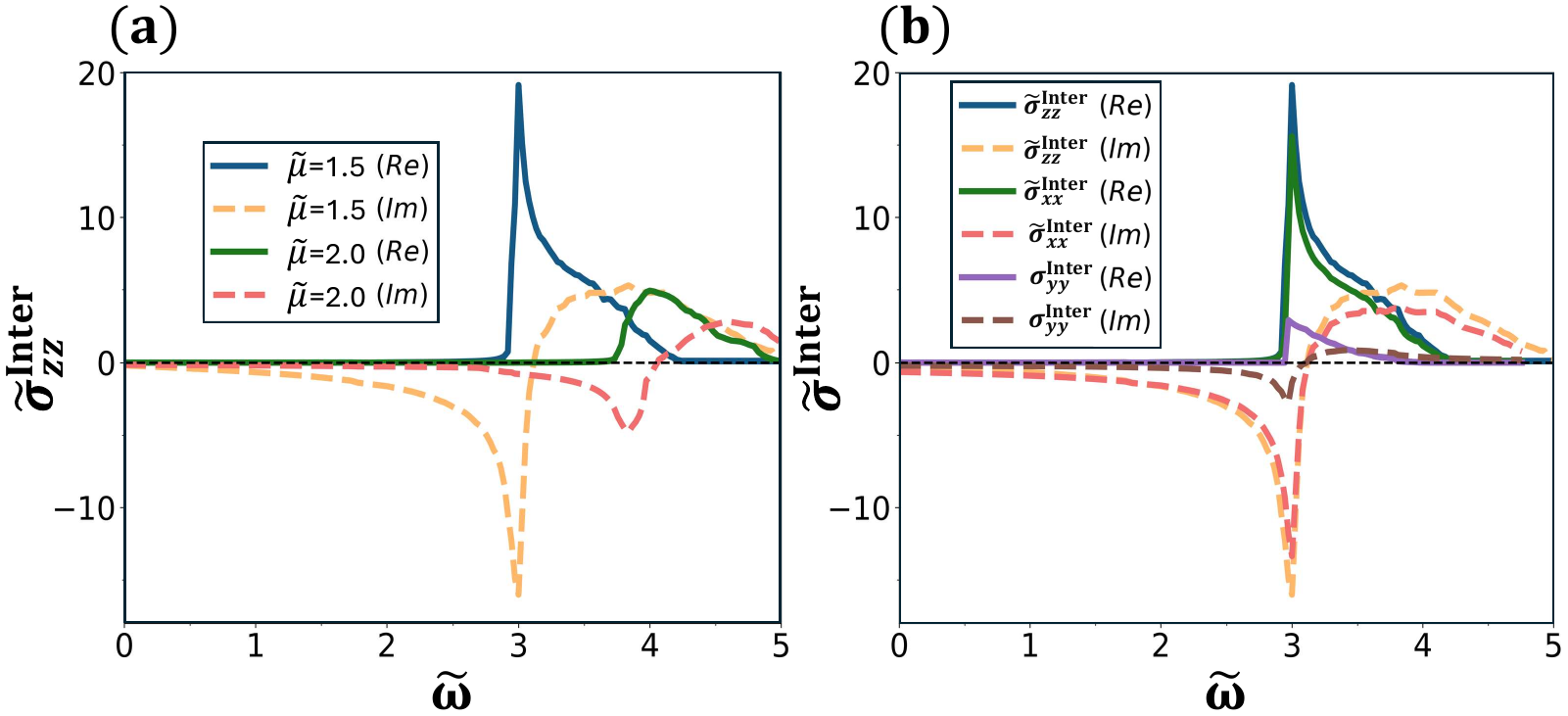}
    \caption{(a) shows the real part (solid line) and imaginary part (dotted line) of the total interband part of longitudinal conductivity due to the field along $z$ direction $\tilde{\sigma}_{zz}^{\text{Inter}}=\sigma_{zz}^{\text{Inter}}/\sigma_0$ where $\sigma_0=e^2/\hbar(2\pi)^2$ as a function of frequency ($\tilde{\omega}$) at distinct values of chemical potential ($\tilde{\mu}$), (b) depicts the comparison of different components of the total interband longitudinal response at fixed $\tilde{\mu}=1.5$. These results are obtained from Eq.~\eqref{eqn:total_zz} by using the model Hamiltonian with the DFT parameters relevant for TaAs. Further, we fixed parameter values as $k_0$ = 0.12$\AA^{-1}$ and $m'$ = 0.01$\AA^{-1}$.} 
    \label{fig:Total}
\end{figure*}

\subsection{Variation with material parameters, $k_0$ and $m'$}
\label{Subsec:internalparameters}

As discussed in Sec.~\ref{subsec:model}, the presence of SOC fundamentally alters the band topology of the nodal line semimetal, driving the system through distinct phases depending on the competition between the nodal ring radius $k_0$ and the SOC related parameter $m'$. Specifically, varying the ratio $k_0/m'$ enables a continuous evolution from a nodal-line semimetal to a Weyl-like phase and to a fully gapped insulating phase. This tunability provides an ideal platform to investigate how changes in band topology and gap formation influence intrinsic and extrinsic interband transport. 
Fig.~\ref{fig:Int_Ext_m_mp}(a) shows the real (solid lines) and imaginary (dotted lines) parts of the intrinsic interband conductivity as a function of frequency at fixed chemical potential for different regimes of $k_0$ and $m'$. 
In the regime $k_0 > m'$, the system behaves as a Weyl semimetal having dispersion as depicted in Fig.~\ref{fig:3D_dispersion}(b), and the intrinsic response exhibits a transition peak at $\tilde{\omega}=2\tilde{\mu}$, corresponding to the onset of interband transitions across linearly dispersing bands. As the system approaches the critical point $k_0 = m'$, SOC opens a finite gap in the dispersion shown in Fig.~\ref{fig:3D_dispersion}(c), which suppresses the magnitude of the intrinsic response while preserving its overall qualitative behavior. This suppression reflects the reduced interband coherence caused by the SOC-induced gap, which increases the energy cost for virtual interband transitions. A similar reduction persists in the gapped regime $k_0 < m'$. 
The corresponding extrinsic interband conductivity is shown in Fig.~\ref{fig:Int_Ext_m_mp}(b), for the same parameter regimes. In the Weyl-like phase ($k_0 > m'$), the extrinsic response closely follows the intrinsic behavior and is mainly led by scattering-driven interband transitions at the Fermi surface. 
However, in contrast to the intrinsic contribution, the extrinsic response shows a strong sensitivity to the material led parameters. At the critical point $k_0 = m'$, where a gap opens, the location of the transition peaks shifts towards the lower frequencies. This shift becomes more pronounced in the regime $k_0 < m'$, where the system remains gapped, and the characteristic transition frequency continues to move to lower $\tilde{\omega}$. 
The imaginary parts of both intrinsic and extrinsic conductivities exhibit trends consistent with their respective real parts, including peak structures and frequency shifts, in accordance with Kramers–Kronig relations.
\begin{table*}[t]
    \centering
    \fontsize{10pt}{10pt}\selectfont  
    \renewcommand{\arraystretch}{1.8}  
    \caption{The table shows the numerical values of intrinsic, extrinsic and total interband conductivity for the nodal line materials listed in the table.}
    \label{tab:SP}
    \begin{center}
    \begin{tabular}{cccccc}
        \toprule
        \makecell{Materials}
        & \makecell{Chemical\\Potential}
        & \makecell{Frequency}
        & \makecell{$\sigma^{\text{Int}}_{zz}$}
        & \makecell{$\sigma^{\text{Ext}}_{zz}$}
        & \makecell{$\sigma^{\text{Inter}}_{zz}$} \\

        \makecell{}
        & \makecell{$\mu$ in meV}
        & \makecell{$\omega$ in meV}
        & \makecell{in units of\\$e^2/\hbar(2\pi)^2$}
        & \makecell{in units of\\$e^2/\hbar(2\pi)^2$}
        & \makecell{in units of\\$e^2/\hbar(2\pi)^2$} \\
        \midrule

        TaAs & 81 & 164 & 4.13 & 15.00 & 19.13 \\
        TaP  & 81 & 164 & 4.51 & 8.11  & 12.62 \\
        TaAs & 81 & 200 & 0.90 & 3.42  & 4.32 \\
        TaP  & 81 & 200 & 1.85 & 3.63  & 5.48 \\

        \bottomrule
    \end{tabular}
    \end{center}
\end{table*}
\subsection{Variation with external parameters, $\tilde{\mu}$ and $\tilde{\omega}$}
\label{Subsec:externalparameters}

In Fig.~\ref{fig:Total}, we show the variation of the total interband part of conductivity for the NLSM system in the presence of SOC with respect to the frequency ($\tilde{\omega}$) at distinct chemical potential ($\tilde{\mu}$) values. Here, the solid curves refer to the real part and the dashed curves refer to the imaginary part of the interband conductivity.

In Fig.~\ref{fig:Total}(a), we observe that the interband conductivity $\tilde{\sigma}^{\text{Inter}}_{zz}$ for NLSMs shows a characteristic transition peak at $\tilde{\omega}=2\tilde{\mu}$ arising as the chemical potential touches the bottom of the conduction band. 
The observed peak corresponds to the onset of Pauli-unblocked interband transitions.  As evident from Figs.~\ref {fig:1D_dispersion} and~\ref{fig:3D_dispersion}, the presence of SOC breaks the SU(2) and time-reversal symmetry~\cite{Fang_cpb2016, Burkov_prb2018}. It thus generates a gap, which affects the band dispersion, making it spin non-degenerate and thereby impacting the total conductivity.  
Further, the position and magnitude of the observed transition peaks can be efficiently tuned by varying the chemical potential. An increase in $\tilde{\mu}$ shifts the characteristic transition frequency toward higher $\tilde{\omega}$, while simultaneously reducing the magnitude of the peak.
The imaginary part of the total conductivity follows the same qualitative behavior as the real part, with a notable sign change occurring at $\tilde{\omega}=2\tilde{\mu}$. This sign reversal signals a resonance-like transition and is consistent with causality constraints imposed by the Kramers–Kronig relations. At very low frequencies, the interband contribution becomes less relevant due to the constraint $\omega\tau\gg1$, where the intraband Drude response dominates. Therefore, the interband results presented here are most meaningful in the finite-frequency regime. 

Fig.~\ref{fig:Total}(b), illustrates the distinct components of the total interband conductivity. Here, the total conductivity $\tilde{\sigma}_{ii}^{\text{Inter}}$ with $i=x,z$ shows nearly the same qualitative and quantitative behavior. However, the strength of $\sigma_{yy}^{\text{Inter}}$ is weak in comparison to other longitudinal components. The reason is the finite nature of $\sigma_{yy}^{\text{Inter}}$ solely due to the field-driven response, while the corresponding scattering-driven contribution vanishes. 
Hence, this shows the directional anisotropy in the total conductivity. In contrast to the longitudinal parts, the anomalous components of total conductivity $\sigma^{\text{Inter}}_{ij} = 0$ vanishes, indicating that symmetry constraints prohibit transverse interband responses in the present model~\cite{Wang_prb2021, Burkov_prb2018}. 

 In conclusion, the intrinsic and extrinsic interband response shows a resonance peak with external parameters. However, with material led parameters, intrinsic response only shows a quantitative suppression with varying $k_0$ and $m'$, its qualitative behavior remains largely unchanged across different topological regimes. On the other hand, the extrinsic interband conductivity accommodates a highly tunable transition peak originating from the Fermi surface whose position and magnitude can be efficiently controlled by the material parameters. 
This difference underscores the critical role of disorder-driven processes in shaping the experimentally observable interband conductivity and highlights the potential of SOC-engineered nodal-line semimetals for tunable optoelectronic and spintronic applications.

\subsection{Experimental relevance and validity}
Experimentally, the longitudinal conductivity of an NLSM can be probed by measuring the transport current under an applied electric bias. A three-dimensional sample is taken and connected to a voltage source, with a bias applied either along the $x$, $y$ or $z$-direction. The resultant current is then measured along the same direction to extract the corresponding longitudinal conductivity components, $\tilde{\sigma}_{xx}$, $\tilde{\sigma}_{yy}$ and $\tilde{\sigma}_{zz}$, respectively. 
To distinguish between intrinsic and extrinsic interband contributions, the chemical potential can be modulated via gating techniques or electronic doping. 
Quantitatively, for an incident field with frequency $\omega \approx 35-40$~THz (corresponding to $\omega\approx140 - 164$~meV), which lies in the mid-infrared/upper terahertz regime, nodal line semimetals having nodal ring radius ranges from $k_0\approx 0.11-0.12\,\AA^{-1}$, based on our results, the interband transitions can be observed at the dimensionless transition frequency $\tilde{\omega}=2\tilde{\mu}=3.0$, corresponds to the chemical potential range $\mu\approx70 - 82$~meV.
The consideration of the low-energy model Hamiltonian helps us to capture the important features of NLSMs in the vicinity of the nodal ring. This model explains the underlying physics of the system near the Fermi level. In this approximation, the transport response is dictated by the inclusion of the nodal line geometry, SOC-induced band splitting, and Berry connection structure. To see the effects of other bands away in the momentum space, one has to go with $\textit{ab initio approach}$ which captures the electronic structure over the entire Brillouin zone.
This can add additional contributions to extrinsic conductivity through the scattering driven matrix elements $U_{\bm k, \bm {k}'}^{np}= \bra{u^n_{\bm k}}U(\bm r)\ket{u^p_{\bm k'}}$ and in the intrinsic conductivity through the Berry connection $\mathcal{R}^{np}_{i}=\langle u^n_{\bm k}|i\partial_{\bm k_i}u^p_{\bm k}\rangle$. Furthermore, the features such as the spin non-degeneracy due to the SOC, the direction anisotropy observed in the interband conductivity, and the tunable interband resonance peak through the disorder and field-driven mechanism remain qualitatively the same even beyond the low energy approximation. However, there can be quantitative modifications in terms of the magnitude and position of the observed resonance peak.

Further, the present study is valid for weak disorder, i.e., $\mu \gg \hbar/\tau$ or $k_Fl\gg1$, where $\tau$ is the relaxation time, $k_F$ represents the Fermi wave vector, and $l$ is the mean free path. The disorder is treated here within the first-order Born approximation using the randomly uncorrelated impurity scattering.
%
%
On considering the interband relaxation time $\tau \approx 10^{-12}$s and $\hbar = 6.6 \times 10^{-13}$ meV$\cdot$s, we find $\mu \gg 0.66$ meV. In the scaled form, our results are valid for $\tilde{\mu} \gg 0.012$. In this regime, the perturbative theory assumes that weak disorder introduces a small modification to the total Hamiltonian, which governs the energy distribution. 
Within this approach, one can keep finite order terms based on the strength of the disorder and can use this approach to understand how disorder influences electronic properties, such as conductivity. Specifically, the approach is effective when the disorder is weak to moderate, but as the disorder intensifies, the first-order Born approximation does not remain valid. 

In contrast, the strong disorder limit, i.e., $\mu \ll \hbar/\tau$, breaks the Bloch band structure, and the topological features becomes non-sharply defined. Hence one has to go beyond the first-order Born approximation, where, the density matrix has to be renormalized by using the spectral function approach. 
Additionally, the whole analysis in this study is performed within the low temperature limit. At high temperatures, the scattering events such as electron-phonon scattering become a key part, thereby modifying the total response significantly, which is beyond the scope of the present work.

\subsection{Numerical analysis}
The numerical comparison of the intrinsic, extrinsic and total interband contribution to the conductivity for TaAs and TaP is summarized in Table~\ref{tab:SP}.
Here, we have taken the numerical value near the interband transition point. All results presented in this work are obtained using DFT-fitted parameters for TaAs~\cite{Pandey_jpcm2023}, which serves as the reference system. From the SOC-induced band gap of 162 meV in TaAs~\cite{Pandey_jpcm2023}, we have calculated the nodal ring radius $k_0=0.12\AA^{-1}$, energy associated with the nodal ring $\varepsilon_0=54.86$ meV and the dimensionless terms $\gamma, \gamma'=1.438$. At $\tilde{\mu}=\mu/\varepsilon_0=1.5$ and $\tilde{\omega}=\omega/\varepsilon_0 = 3.0$, the intrinsic conductivity is $\sigma^{\text{Int}}=4.13$, whereas the extrinsic conductivity gives the value $\sigma^{\text{Ext}}=15$, so $\sigma^{\text{Ext}}=3.5\sigma^{\text{Int}}$, yielding the total conductivity $\sigma^{\text{Inter}}=19.13$. Here, the conductivities are expressed in units of $e^2/\hbar (2\pi)^2$.\\
\section{Summary}\label{Sec:summary}
In this work, we have systematically investigated the interband conductivity of NLSMs in the presence of SOC using a quantum kinetic approach. The interband response is shown to arise from the field-assisted intrinsic process and the scattering-driven extrinsic mechanism. We find that the interband response shows a prominent characteristic transition peak when the chemical potential touches the edge of the conduction band, reflecting the onset of Pauli-unblocked interband transitions.

We further highlight that this interband peak is highly sensitive to the system's external parameters, such as $\tilde{\mu}$ and $\tilde{\omega}$. Additionally, the response shows strong tunability with the material parameters, namely the nodal ring radius $k_0$ and the SOC-related parameter $m'$. While the intrinsic response only shows a quantitative change, the extrinsic response shows a tunable feature across different topological regimes. The resulting interband conductivity also exhibits a directional anisotropy, providing a distinctive fingerprint for experimental detection.

To establish a direct connection with the real materials, our results are obtained by considering the DFT fitted parameters for TaAs. However, the observed features are generic to 3D system, irrespective of the choice of material.
Similar behavior is expected in other materials, such as NbAs, NbP, and PbTaSe$_2$. In these materials, while qualitative behavior, such as the emergence of interband resonance peaks and anisotropic conductivity, remains the same, the exact location of resonance and the magnitude of the interband conductivity can change depending on the material band topology.

The tunable interband response, combined with SOC, lifts spin degeneracy and topological phase transitions, showing the potential application of these NLSM materials in spintronic devices, optoelectronic applications, and topological transistors~\cite{Chi_am2018, Garcia_prr2020, Nourizadeh_aqt2023, Zhong_am2025,shi_apl2015}.
%


%
%
%

\section{Acknowledgment}
We acknowledge D. Joy for insightful discussion and valuable suggestions. This work is financially supported by Anusandhan National Research Foundation-under project number SUR/2022/000289.

\onecolumngrid
\appendix
\section{Derivation of Equation(4)} \label{Appendix:A}
Following Eq.~\eqref{eqn:QLE} for $S_{\bm E,{\bm k}}^{np}$ we can write
\begin{equation} \label{eqn:A6}
    \frac{\partial  S_{\bm E,{\bm{k}}}^{np}}{\partial t}+\frac{i}{\hbar} [ \mathcal{H}_0, S_{\bm E,{\bm{k}}}]^{np}+\mathcal{J}_{\bm{k}}^{np}[\mathcal{N}_{\bm E,{\bm k}}] + \mathcal{J}_{\bm{k}}^{np}[\mathcal{S}_{\bm E,{\bm k}}]={D_{\bm E,{\bm{k}}}^{np}},
\end{equation}
here the term  $\langle n| [ \mathcal{H}_0, S_{\bm E,{\bm{k}}}]|p\rangle= S^{np}_{\bm E,{\bm{k}}}(\varepsilon_{\bm k}^{n}-\varepsilon_{\bm k}^{p}) $, where  $\mathcal{H}_0$ is the band Hamiltonian and $S_{\bm E, \bm k}$ is the filed dependent interband part of density. Here, $D_{\bm E,{\bm{k}}}^{np}$ represents the field assisted term and $\mathcal{J}_{\bm{k}}^{np}[\rho_{\bm E,{\bm k}}] = \mathcal{J}_{\bm{k}}^{np}[\mathcal{N}_{\bm E,{\bm k}}]+\mathcal{J}_{\bm{k}}^{np}[\mathcal{S}_{\bm E,{\bm k}}]$ represents the scattering driven term, where $\mathcal{J}_{\bm{k}}^{np}[\mathcal{N}_{\bm E,{\bm k}}]$ and $\mathcal{J}_{\bm{k}}^{np}[\mathcal{S}_{\bm E,{\bm k}}]$ take care of the intraband and interband contribution to the field driven scattering term, respectively. Further, we have evaluated the $\mathcal{J}_{\bm{k}}^{np}[\mathcal{N}_{\bm E,{\bm k}}]$ by using the first order Born approximation, however, the $\mathcal{J}_{\bm{k}}^{np}[\mathcal{S}_{\bm E,{\bm k}}]$ is handled within the relaxation time approximation.
Further, we can write
\begin{equation} 
    \frac{\partial  S_{\bm E,{\bm{k}}}^{np}}{\partial t}+\frac{S^{np}_{\bm E,{\bm{k}}}}{\hbar}\big[i \omega^{np} + \eta\big]={D_{\bm E ,{\bm{k}}}^{np}}-\mathcal{J}_{\bm{k}}^{np}[\mathcal{N}_{\bm E,{\bm k}}].
\end{equation}
Here, $\omega^{np}=\varepsilon_{\bm k}^{n}-\varepsilon_{\bm k}^{p} $, $\eta\to0^+$ infinitesimal value. The solution of above differential equation becomes, 
\begin{align}\nonumber
S_{\bm E,\bm{k}}^{np}=\frac{1}{\hbar^2}\int_0^\infty dt' e^{-i\mathcal \omega^{np} t'/\hbar} \space \big({D_{\bm E,{\bm{k}}}^{np}} -\mathcal{J}_{\bm{k}}^{np}[\mathcal{N}_{\bm E,{\bm k}}]  \big)e^{i\mathcal \omega^{np} t'/\hbar}.
\end{align} 
Further, the solution of the interband density matrix in the presence of oscillating electric field $\bm E (t)= \bm E e^{-i \omega t}$ can be written in the combination of the field driven and scattering driven parts, 
\begin{align}\nonumber
S_{\bm E,\bm{k}}^{np}=\frac{{D_{\bm E,{\bm{k}}}^{np}} -\mathcal{J}_{\bm{k}}^{np}[\mathcal{N}_{\bm E,{\bm k}}] }{\eta+i \big( \omega^{np}+ \hbar \omega\big)}.
\end{align} 
%
Conductivity can be expressed as $\bm j =\sigma \bm E= \text{Tr}[{\bm v}\rho]= \sum_{\bm k}v \rho$, here $\bm j$ represents the current density and $\sigma$ is the symbol used for conductivity. 
The $\bm v$ show velocity of the electron in band basis representation and in normalized unit $ \space \tilde{v}^{pn}_i=\space\delta_{pn}\space\partial_{\tilde{k}_i}\tilde{\varepsilon}_{\bm{{k}},\beta}^n+i\space\mathcal{\tilde{R}}^{pn}_{i}\space\tilde{\omega}^{pn}$, where $\tilde{v}_i=\hbar  {k}_0 v_i/\varepsilon_0 $, $\tilde{\omega}^ {pn}= \omega^{pn}/\varepsilon_0$ and $\mathcal{\tilde{R}}_i=\mathcal{\tilde{R}}_{k_i}$.
Further, the interband part of the conductivity can be written as $\sigma^{\text{Inter}}= \sum_{\bm k}\tilde{v}^{pn}_i S^{np}_{\bm E, \bm k}$, keeping the values of interband velocity and density matrix we get
\begin{align}
    \sigma^{\text{Inter}}= \frac{i e}{\bm E} \sum_{\bm k}\space\mathcal{\tilde{R}}^{pn}_{i}\space\tilde{\omega}^{pn} \Bigg\{\frac{{D_{\bm E,{\bm{k}}}^{np}} -\mathcal{J}_{\bm{k}}^{np}[\mathcal{N}_{\bm E,{\bm k}}] }{\eta+\space i \space ( \tilde{\omega}^{np} + \tilde{\omega})}\Bigg\}
\end{align}
Here, $ D_{\bm E,\bm{k}}^{np} = \langle n|[\mathcal{H}_{\bm E},\rho]\ket{p} = e\bm{E}/\hbar.\left(\partial_{\bm {k}} \rho^{np}-i \mathcal{R}_{\bm{k}}^{np} F^{np}\right)$~\cite{Bhalla_prb2023, Nagaosa_AM2017}, the $\mathcal{R}_{\bm{k}}^{np}=\langle{u_{\bm{k}}^n}\ket{i\partial_{\bm{k}}u_{\bm{k}}^p}$ is Berry connection in the $\bm{k}$ space and $|u_{{\bm k}}^{n} \rangle$ represents the eigenfunction coming from the system's Hamiltonian.
The final form of the interband conductivity in normalized unit becomes

\begin{align}
\sigma_{ii}^{\text{Inter}}=e \sum_{n\neq p}\sum_{{\bm k}}  \frac{ \mathcal{\tilde{R}}_{i}^{pn} \tilde{\omega}^{pn}}{\eta+\space i \space ( \tilde{\omega}^{np} + \tilde{\omega})} \Big(\frac{e \mathcal{\tilde{R}}_{i}^{np}\tilde{F}^{np}}{\hbar}  - \frac{i\mathcal{J}_{\tilde{{k}}_i}^{np}[\mathcal{N}_{\bm E}]}{E_i}\Big).
\end{align}
Here, this is the same Eq.~\eqref{eqn:total_zz} use to calculate the interband conductivity of the NLSMs.
Further, form of the scattering term in the weak disorder limit within the first order Born approximation becomes,
\begin{align}\nonumber
\mathcal{J}_{\bm{k}}^{np}[\mathcal{N}_{\bm E, \bm k}]=\frac{1}{\hbar^2}\int_0^\infty dt' \sum \big[\langle U_{\bm{kk}'}^{nq}U_{\bm{k}'\bm{k}}^{qn}\rangle\space\mathcal{N}_{\bm E,\bm{k}}^{qp}e^{i\big(\varepsilon_{\bm{k}'}^q-\varepsilon_{\bm{k}}^n\big)t'/\hbar}-\langle U_{\bm{kk}'}^{nq}U_{\bm{k}'\bm{k}}^{np} \rangle\space\mathcal{N}_{\bm E,\bm{k}'}^{qn}e^{i\big(\varepsilon_{\bm{k}'}^n-\varepsilon_{\bm{k}}^p\big)t'/\hbar}\\
-\langle U_{\bm{kk}'}^{nq}U_{\bm{k}'\bm{k}}^{np}\rangle\space\mathcal{N}^{qn}_{\bm E,\bm{k}'}e^{i\big(\varepsilon_{\bm{k}'}^n-\varepsilon_{\bm{k}}^q\big)t'/\hbar}+ \langle U_{\bm{kk}'}^{qn}U_{\bm{k}'\bm{k}}^{np}\rangle\space\mathcal{N}^{nq}_{\bm E,\bm{k}}e^{i\big(\varepsilon_{\bm{k}'}^q-\varepsilon_{\bm{k}}^n\big)t'/\hbar}\big],
\end{align} 
where $\langle U_{\bm{kk}'}^{nq} U_{\bm{k}'\bm{k}}^{qt}\rangle = U_0^2 \langle u_{\bm{k}}^n | u_{\bm{k}'}^q \rangle \langle u_{\bm{k}'}^q | u_{\bm{k}}^t \rangle$.

\twocolumngrid
\bibliography{Ref}

\end{document}